# Impurity concentration dependent electrical conduction in germanium crystal at low temperatures

Manoranjan Ghosh,[*] Shreyas Pitale, S.G. Singh, Shashwati Sen, S.C. Gadkari

*Crystal Technology Section, Technical Physics Division,*
*Bhabha Atomic research Centre, Trombay, Mumbai – 400088*
[*]Email: mghosh@barc.gov.in

**Abstract:** Germanium single crystal having 45 mm diameter and 100 mm length of 7N+ purity has been grown by Czochralski method. Structural quality of the crystal has been characterized by Laue diffraction. Electrical conduction and Hall measurements are carried out on samples retrieved from different parts of the crystal along the growth axis. Top part of the crystal exhibits lowest impurity concentration ($\sim 10^{12}/cm^3$) that gradually increases towards the bottom ($10^{13}/cm^3$). The crystal is n-type at room temperature and the resistivity shows non-monotonic temperature dependence. There is a transition from n-type to p-type conductivity below room temperature at which bulk resistivity shows maximum and dip in carrier mobility. This intrinsic to extrinsic transition regions shift towards room temperature as the impurity concentration increases and reflects the purity level of the crystal. Similar trend is observed in boron implanted high purity germanium (HPGe) crystal at different doping level. The phenomena can be understood as a result of interplay between temperature dependent conduction mechanism driven by impurity band and intrinsic carrier in Ge crystals having fairly low acceptor concentrations ($<10^{12}/cm^3$).

# 1. Introduction

Germanium is a group VI indirect semiconductor like silicon but with a smaller band gap and higher carrier mobility [1,2]. The purest form of germanium shows high resistivity and high electron and hole mobility at low temperature. These properties make Ge more attractive than Si for opto-electronic applications. However, presence of unstable oxide layer at the surface restricts its application in electronic devices. Relatively low purity crystals find applications in wave guide, IR window fiber optics, infrared night vision devices, space solar cells and as polymerization catalyst [3]. The high purity Ge crystals (HPGe) are used for fabrication of thick gamma ray detector [4,5]. Therefore, growth of pure germanium crystal is highly desirable especially for nuclear detector application [6-8]. Single crystals of germanium are grown at various purity levels. Particularly, purification of raw material and growth of HPGe is remained a challenging task to many laboratories [9-12]. The impurity present therein (concentration~$10^{10}$/cm$^3$) are primarily un-intentional. Doping of selective dopants is also performed for the fabrication of detector and other electronic devices [13,14]. Therefore, it is important to understand the electrical conduction properties of Ge at various purity levels.

Temperature dependent transport properties of Germanium are highly sensitive to its purity level [15]. Germanium having impurity concentration above $10^{15}$/cm$^3$, shows anomalies in resistivity, hall co-efficient and mobility at low temperature [16]. It is proposed that impurity band conduction is primarily responsible for such behaviour [17]. There are limited studies on conduction properties of Ge with impurity concentration ranging from $10^{10}$ to $10^{15}$/cm$^3$. Also the type of conductivity (impurity) has visible impact on the conduction properties. In few studies with higher level of impurity concentration, Ge crystals do not exhibit change in conductivity throughout the temperature range [16]. In some studies crystal with lower impurity concentration shows n-type at higher temperature and p-type at low temperature [18]. Resistivity and mobility curve also shows different behaviour when such transition occurs. Therefore exhaustive studies on transport properties of Ge with lower level of impurity concentration are relevant.

In this study, 7N pure Ge crystal is grown and characterized. Transport properties of Ge crystal with impurity concentration in the range of $10^{10}$ to $10^{14}$/cm$^3$ have been investigated. p-type Ge crystals of such purity change its type of conductivity with associated changes in resistivity and mobility at certain temperature. It is observed that the transition in type of conductivity occurs at higher temperature when impurity concentration increases.

## 2. Growth and processing of Ge crystal

Germanium single crystal of 45 mm diameter and 100 mm length of 7N+ purity has been grown by Czochralski method (figure 1a) from a zone refined polycrystalline ingot (weight 850 gm) supplied by the Chemistry Division, BARC. The charge consisting of zone refined ingot is etched in concentrated $HNO_3$ (65%):HF(48%)::3:1 solution for 10 minutes, rinsed under running DI water (>18 MΩ) and dried under high pure $N_2$. Then the material is placed in a quartz crucible (cleaned by dipping in 10% HF and rinsing in DI $H_2O$) of 100 mm diameter and 150 mm length. Seed crystal (oriented towards <100> direction) were cleaned and etched by following similar steps as mentioned above and used for the crystal growth. A graphite susceptor is coupled with a 50 kW, 10-13 Hz RF generator to melt the material. However, it is found that at high temperature germanium gets coupled with the RF thus making the seeding procedure difficult. This problem was avoided by adjusting the melt height. A typical rotation rate of 15-20 rpm and a pull rate of 20-25 mm/h are used during the growth of Ge single crystal. The growth experiments are carried out in class-10000 clean room while material preparation is done in class-1000 clean environment. The purity of grown crystal is comparable to that of the initial charge (polycrystalline zone refined ingot) as verified by analytical techniques (data not shown). Thus, no extra impurity was added during the crystal growth process.

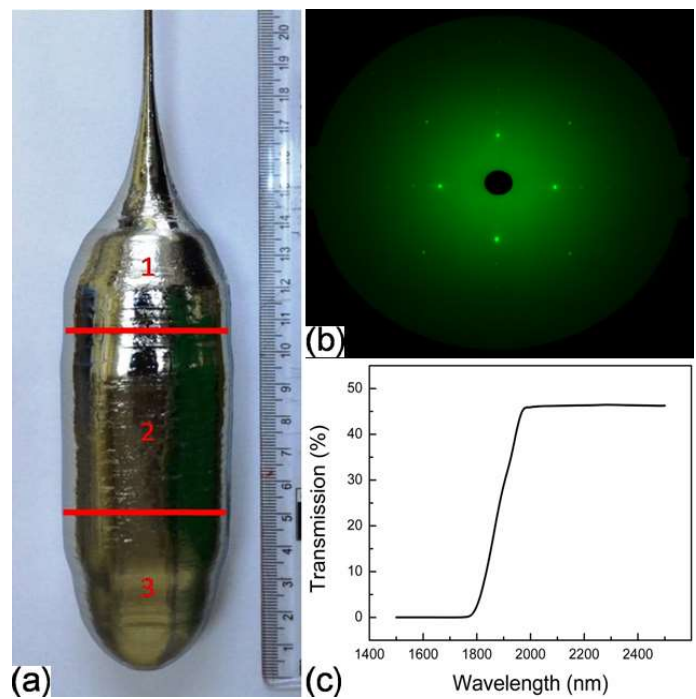

**Figure 1.** (a) 7N pure as grown germanium single crystal (b) Laue diffraction pattern of grown crystal (c) Transmission spectra for a 5 mm thick germanium cut from the grown crystal.

The grown crystal was cut into slices of desirable size by a diamond wire saw cutting machine. Then the slices are further cut into pieces of intended size and shape. All the pieces were subsequently ground and lapped. Lapping is performed in steps by silicon carbide abrasive paper with different grit sizes (600-1500). All pieces are polished by diamond paste 10 μm, 5 μm, 3 μm and 0.3 μm particle sizes to remove lapping imperfections and to achieve smooth mirror finish surface. Crystals are further cleaned using ultra-sonication in suitable solvents to remove organic and inorganic impurities from the surface. Finally the crystal slices are etched before formation of electrical contacts. The crystals are stored in an inert atmosphere glove box for further use.

Structural characterization of the crystal is carried out by Laue back-reflection and transmission measurements (1500-2500 nm). For this purpose, the grown crystal is cut in to 5 mm thickness and with varying cross sections (5x5 mm$^2$ for electrical and structural measurement and 35x35 mm$^2$ for optical measurement). The Laue back-reflection pattern (recorded using Bruker make Laue camera) of the grown crystal is shown in figure 1b. The <100> orientation of the grown crystal was confirmed by fitting the back-reflection pattern using Orient-express code. The transmission spectra recorded (by Shimadzu 3600 spectrophotometer) for the Ge sample (35x35x5 mm$^3$) is shown in figure 1c. Optical band gap calculated from the transmission data is found to be ≈0.70 eV.

## 3. Minority carrier lifetime measurement

The residual impurity concentration in HPGe crystal is beyond the detection limit of standard analytical techniques. Physical characterization techniques like Hall measurement and Minority Carrier Lifetime is carried out to understand transport and recombination of carriers in Ge crystals.

Life time measurements are found useful for determining the presence of special types of crystal imperfections, which are usually present in miniscule amounts that cannot be detected by Hall or resistivity measurements [19]. The degree of crystal perfection is determined in terms of the maximum value of carrier lifetime [20]. Knowledge of average carrier lifetime gives an estimation of the purity level of crystals grown in this laboratory. The laboratory grown germanium crystal and Umicore HPGe (both p and n-type) crystals are characterized through Minority Carrier Lifetime (MCLT) measurement using Edinburgh FLP 920 instrument and a Keithley current source (Model: 6221). The samples were mounted in an Oxford make optical cryostat and excited using pulsed xenon source at 50Hz frequency.

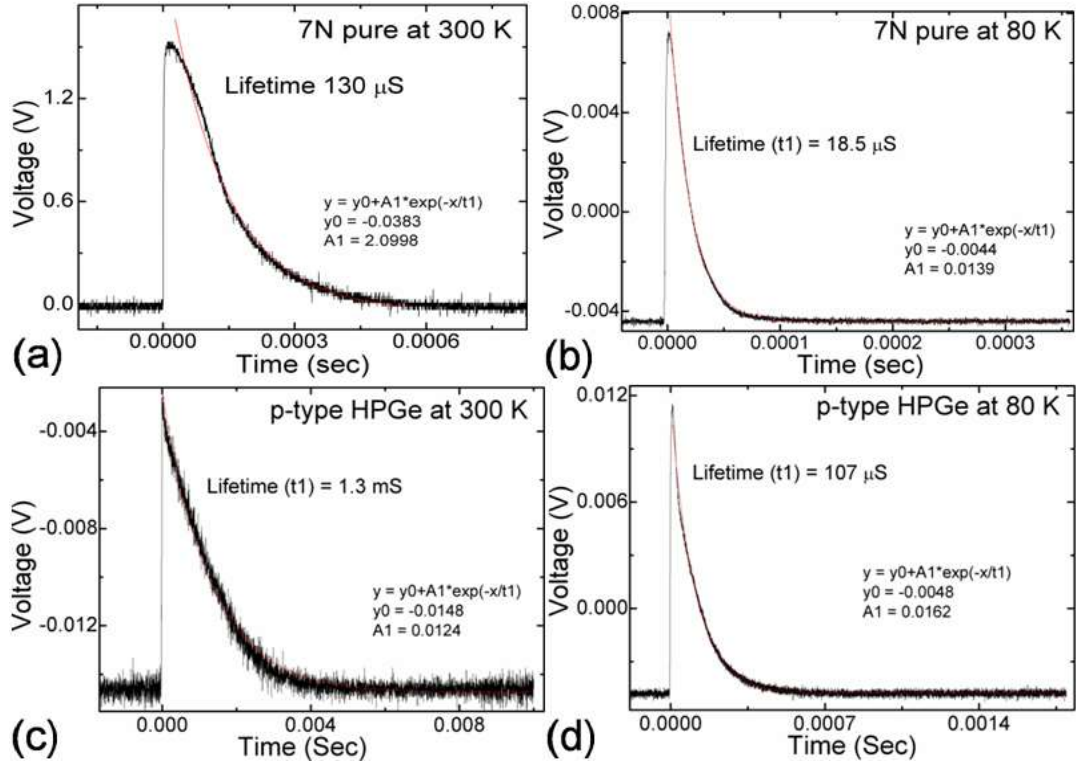

**Figure 2.** Minority carrier lifetimes of grown germanium (a,b) and p–type HPGe (c,d) crystal as indicated on the view graph at 300 and 80 K. Red line indicates single exponential fitting of experimental data.

Signals were recorded on a Tektronix digital oscilloscope. Measurements were performed both at room temperature and 80 K.

The steady-state equilibrium of charge carriers is disturbed and excess minority charge carriers are generated by optical excitation using pulsed xenon lamp. Majority charge carriers are then drawn into the sample from an external current source to preserve space charge neutrality. The resultant effect is the excess density of both charge carriers. After the removal of disturbance, the excess charge carriers establish equilibrium through various processes. Carrier lifetimes are extracted by single exponential fitting of the spectra. The lifetime thus measured for the Umicore HPGe crystal (figure 2c, d) corroborate well with the suppliers data sheet. Usually, the crystal with longer carrier lifetime is considered as better candidate for fabrication of p-i-n diode. The decay curves recorded for 7N pure germanium and HPGe crystals are shown in figure 2a, b. Lifetime recorded at room temperature is nearly 10 times higher than that measured at 77 K. The room temperature lifetime of 7N pure Ge is found to be 130 μSec. On the other hand HPGe crystal shows lifetime > 1 mS.

## 4. Temperature Dependent Hall Measurement by Van der Pauw method

Basic semiconductor properties of Ge are studied by four probe resistivity measurement using Van der Pauw method [21,22]. Parameters such as resistivity, impurity concentration and carrier mobility are determined in this study by temperature dependent Hall measurements conducted using Ecopia make HMS5000 Hall measurement system. Ge slices were cut from three sections of the grown crystal viz. top, middle and bottom part as indicated in figure 1a. Hall measurements was performed according to the procedure described in IEEE standard [23] on samples from each regions having lamellar and square in shapes with shortest linear dimension along a face being at least three times greater than the thickness. Surface roughness and strains (such as those caused by lapping) was removed by polishing and etching. Four contacts were attached on the corners of polish etched square samples at 90° spacing using In-Ga eutectic.

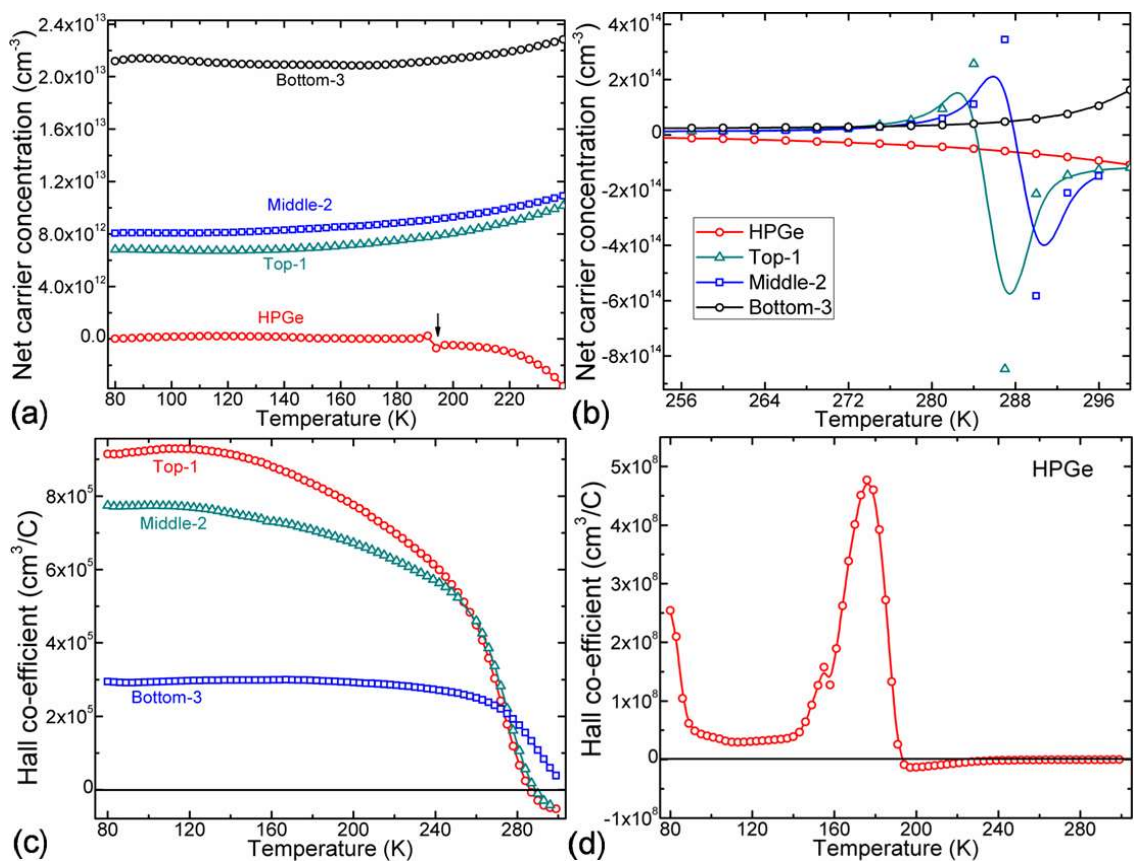

**Figure 3. (a)** Temperature dependent net carrier concentration of HPGe and different parts of 7N pure germanium crystal, (b) expanded view of the same near room temperature. Hall co-efficients of 7N pure germanium (c) and HPGe crystal (d).

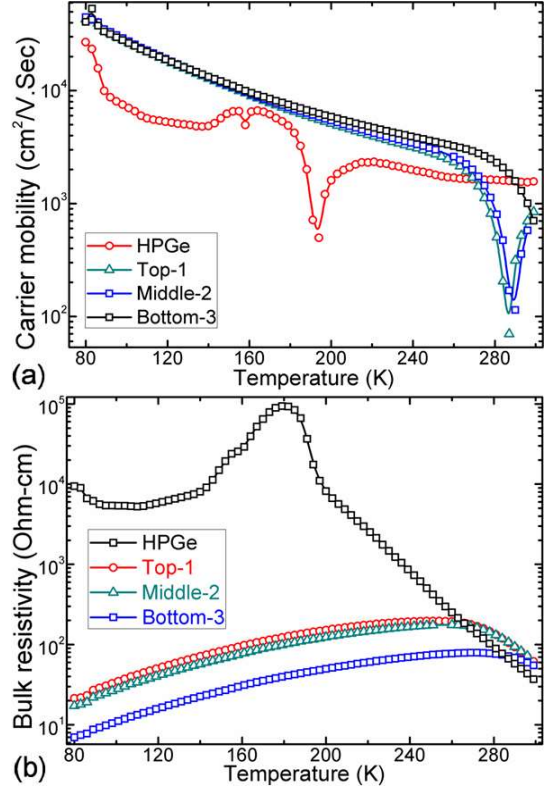

**Figure 4.** Temperature dependent (a) carrier mobility and (b) resistivity of HPGe and different parts of 7N pure germanium crystal.

Net carrier concentration and Hall co-efficient are plotted as a function of temperature in figure 3a-d. HPGe crystal at 77K exhibits p-type conductivity with measured carrier concentration of $8.4 \times 10^9$/cm$^3$. Top part of the 7N pure Ge crystal also shows p-type conductivity with higher carrier concentration of $7.1713 \times 10^{12}$/cm$^3$ at 77 K. Middle and bottom part of the crystals show even higher carrier concentration (up to $2.2 \times 10^{13}$/cm$^3$) as depicted in figure 3a. All crystals show p-type conductivity at low temperature signifying that dopants are acceptor in nature (figure 3c). Carrier concentration increases with the rise in temperature due to creation of thermally generated charge carriers. For HPGe crystal, thermally generated electron starts dominating around 190K and the crystal shows n-type conductivity (figure 3d). Thus extrinsic to intrinsic transition at 190K is associated with the change in type of conductivity of the crystal. 7N pure crystals with higher acceptor concentration exhibit this extrinsic to intrinsic transition at higher temperatures (after 280 K). Top and middle part of the crystal undergoes p to n-type transition at 283 and 290 K respectively. Bottom part with maximum acceptor concentration does not show this transition within 300K. It is evident that crystal with acceptor type impurity range $10^{10}$ to $10^{13}$/cm$^3$ show p-type to n-type transition in conductivity when thermally generated intrinsic electrons dominate the acceptor dopants [24]. If the acceptor dopant concentration is higher than transition occurs at higher temperatures. Due to the interplay between impurity and intrinsic carriers, the resultant mobility shows a steep dip at the point when the transition in type of conductivity occurs (figure 4a). Bulk resistivity also shows maximum at this transition point due to the loss in carrier mobility (figure 4b). For all the samples carrier mobility increases with decreasing temperature and the bulk resistivity first increases and reverses its trend after the type transition by continuing lowering the temperature. N-type crystal or crystal with higher impurity concentration may not show such type transition [25]. Because drift mobility of such Ge crystal does not show dip in temperature dependent profile [26].

**5. Hall measurement of boron ion implanted samples**

Similar phenomena have been observed in intentionally doped HPGe crystal with varying impurity concentration. Intentional doping is performed by boron ion implantation using Danfysik (1080-30) ion implantation system. The ion source essentially consists of the discharge chamber and a directly heated oven which is charged with solid substances containing the material to be ionized ($BN+B_2O_3$ in the present case). A concentrically wound tungsten filament forms the cathode for electrode emission inside discharge chamber. One positively charged anode ring of tungsten with a central bore is used as an inlet aperture for the source material evaporated in the oven. In the discharge space between cathode and anode, ions are formed when electrons are emitted from the cathode, i.e., the tungsten filament. The discharge chamber on the cathode side is linked to a tungsten plate having a narrow central bore of about 0.5 mm diameter. Through this bore the ions are extracted from the ion plasma by means of a highly negatively charged electrode, connected to a potential of at least 10,000 V, and are subsequently fed to the acceleration path and the analyzing magnet, respectively. Once extracted, the ions travel a mass separator where defined m/q is selected by a sectorial magnetic field. An electrostatic scanning system is used to deflect the ion trajectories. This allows uniform irradiation of the sample (better than 3%) with low current density (of the order of 1 $\mu A/cm^2$). The implantation was carried out along <100> direction. The dose is measured by current integrator, which integrates the current obtained from four faraday cups, using following formula:

$$Dose, Q = \frac{\left[\frac{ion\ beam\ current\ in\ amps}{q}\right] \times \left[implant\ time\right]}{\left[implant\ area\right]}$$

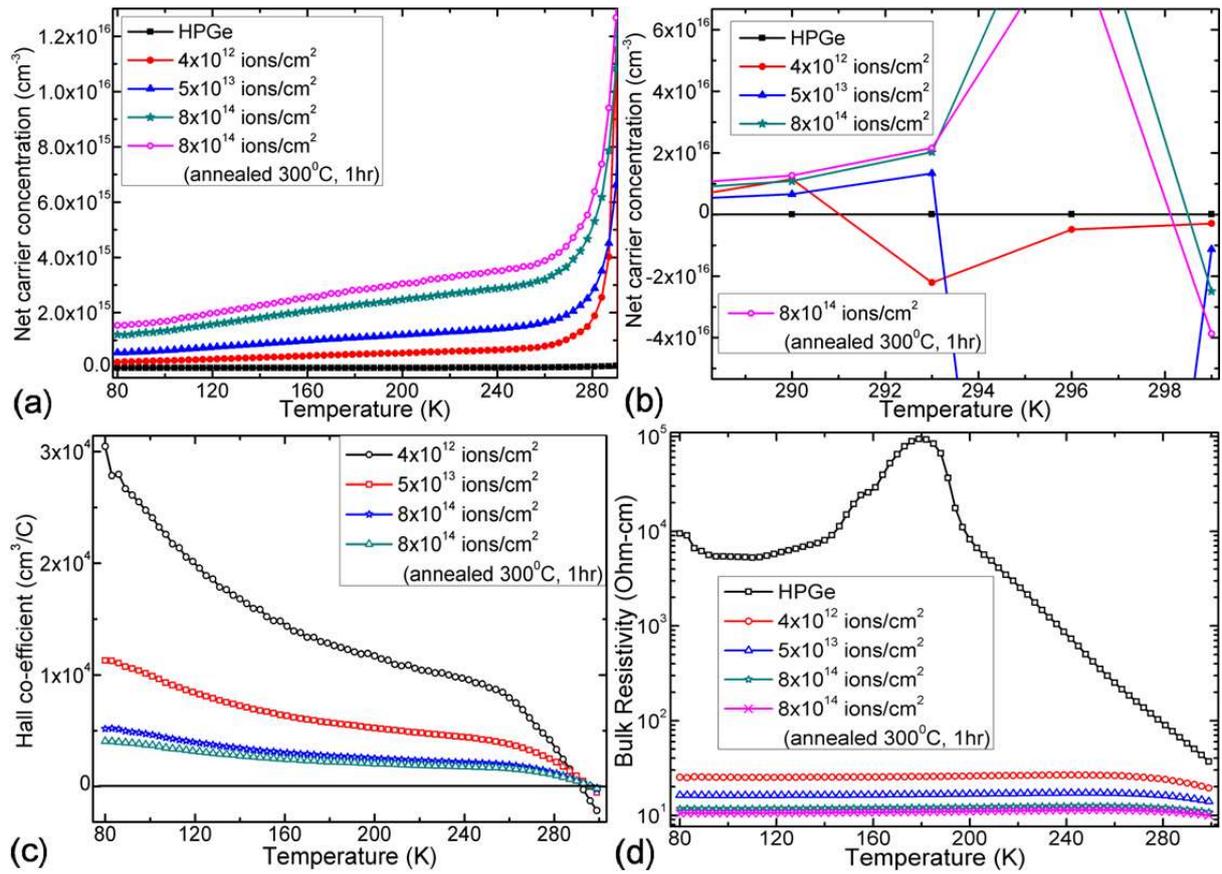

**Figure 5.** (a) and (b) Temperature dependent net carrier concentration, (c) Hall co-efficient and (c) bulk resistivity of boron implanted HPGe crystal with different doses.

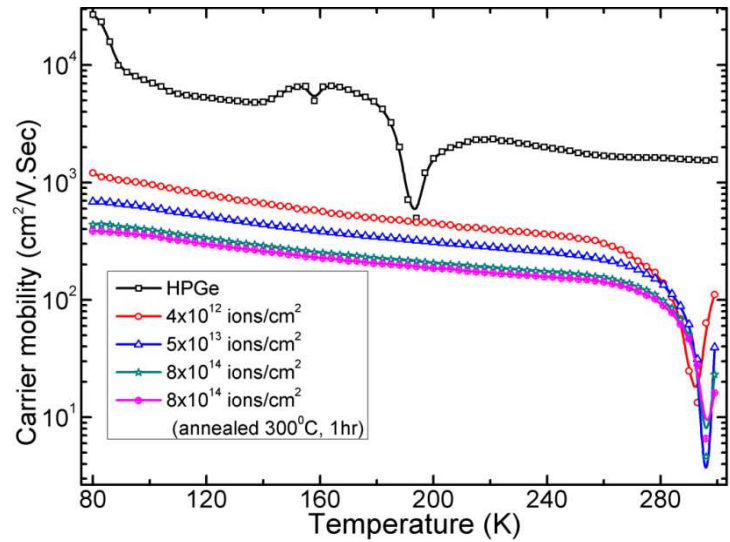

**Figure 6.** Carrier mobility of boron implanted HPGe crystal with different doses.

Temperature dependent (300-80 K) carrier concentration, hall co-efficient and resistivity were obtained on $^{11}$B implanted germanium using Hall measurements in the dose range $10^{12}$-$10^{14}$/cm$^2$ as shown in figure 5. Results for un-implanted HPGe are also included for comparison purpose. We also measured the influence of annealing on the electrical activity of the implanted boron ions. Net carrier concentration of HPGe crystal increases throughout the temperature with increasing doses of implantation (figure 5a). As a result bulk resistivity also reduces for higher boron doses (figure 5d). The carrier concentration is highest (~$10^{17}$/cm$^3$) near room temperature due to supply of additional carriers through thermal generation. At lower temperatures boron doped HPGe show p-type conductivity. At certain temperature thermally generated electron overpopulates the acceptor type boron impurity and the crystal exhibits n-type conductivity (figure 5c). This transition occurs at higher temperatures when the boron impurity concentration increases as seen for 7N pure crystal having different impurity concentration. HPGe crystal with doses nearly $10^{12}$, $10^{13}$, and $10^{14}$ions/cm$^3$ show transition from p-type to n-type conductivity at 290, 294 and 298 K respectively (figure 5b, c). After implantation, annealing at 300$^0$C is performed to increase the activation percentage of dopants [27]. Minor changes in electrical conduction are observed after annealing at 300$^0$C. Thus room temperature electrical activation of boron in germanium is possible at the chosen implantation energy and the annealing step may be avoided to greatly reduce the risk of unwanted diffusion of electrically active metallic impurities. The depth of boron implanted layer in Ge is found to be $\approx$ 0.1µm. As expected, bulk resistivity drastically reduces after boron implantation. As the boron doses increases the resistivity gradually decreases showing nearly10 ohm at 77K for doses 8x$10^{14}$ions/cm$^2$ (figure 5d). Carrier mobility of boron implanted HPGe crystal with different doses also shows similar traits as seen for the 7N pure Ge crystal (figure 6). Due to interplay between impurity and thermally generated carriers, the drift mobility shows a dip near the extrinsic to intrinsic transition region.

## 6. Conclusions

Large size Ge single crystal of 7N purity has been grown. Physical properties of three different parts of the crystal along the axis with varying impurity concentration have been investigated. Top part of the grown crystal exhibits net carrier concentration $\approx 10^{12}$/cm$^3$ that gradually descends to $\approx 10^{13}$/cm$^3$ at the bottom at 80 K. It is found that p-type crystals having carrier concentration below$10^{13}$/cm$^3$ exhibits extrinsic to intrinsic transition near room temperature. This transition is associated with change in type of the crystal from p-type to n-type conductivity. The temperature at which this conductivity 'type' transition occurs increases with the net carrier concentration that can be considered as a mark of impurity index of the crystal. Similar phenomena have been observed in boron implanted HPGe crystal (carrier conc. ~$10^{10}$/cm$^3$) that further validate the findings from 7N pure Ge crystals.

## Acknowledgements

Authors are thankful to all members of Crystal Technology Section for their unconditional help and support.